\documentclass[aps,showpacs]{revtex4}
\usepackage[dvips]{graphicx}
\usepackage{amsmath}

\begin{document}

\title{Ginzburg - Landau equation from SU(2) gauge field theory}
\author{Vladimir Dzhunushaliev}
\email{dzhun@hotmail.kg}
\affiliation{Freie Universit\"at Berlin, Fachbereich Physik, Arnimalleee 14, 
14195 Berlin, Germany \\
and \\
Dept. Phys. and Microel. Engineer., Kyrgyz-Russian
Slavic University, Bishkek, Kievskaya Str. 44, 720000, Kyrgyz
Republic}

\author{Douglas Singleton}
\email{dougs@csufresno.edu}
\affiliation{Physics Dept., CSU Fresno, 2345 East San Ramon Ave.
M/S 37 Fresno, CA 93740-8031, USA}

\date{\today}

\begin{abstract}

The dual superconductor picture of the QCD vacuum is thought to describe
various aspects of the strong interaction including confinement. Ordinary
superconductivity is described by the Ginzburg-Landau (GL) equation. In the present 
work we  show that it is possible to arrive at a GL-like equation
from pure SU(2) gauge theory. This is accomplished by using Abelian projection 
to split the SU(2) gauge fields into an Abelian subgroup and its coset. The 
two gauge field components of the coset part act as the effective, complex, scalar 
field of the GL equation. The Abelian part of the SU(2) gauge field 
is then analogous to the electromagnetic potential in the GL equation.
An important aspect of the dual superconducting model is for the GL 
Lagrangian to have a spontaneous symmetry breaking potential, and the existence
of Nielsen-Olesen flux tube solutions. Both of these
require a tachyonic mass for the effective scalar field. Such a
tachyonic mass term is obtained from the condensation of ghost fields.
\end{abstract}


\maketitle

\section{Introduction}

In ref. \cite{dzhsin02a} it was shown that one could obtain
London's equation from SU(2) gauge theory using the ideas of
Abelian projection. This supports
the dual superconductor picture of a confining Yang-Mills gauge
theory, since London's equation gives a phenomenological 
description of the Meissner effect. The vacuum of the SU(3) Yang-Mills
theory of the strong interaction
is thought to exhibit a dual Meissner effect with respect to
chromoelectric fields in analogy to a normal superconductor, which 
exhibits the Meissner effect with respect to the magnetic field.
In \cite{dzhsin02a} the quantized SU(2) gauge fields split into two phases: 
an ordered and  a disordered phase. The ordered phase was the gauge field 
belonging to  the U(1) subgroup (the diagonal,  Abelian component of 
the SU(2) gauge field) and the disordered phase was the coset SU(2)/SU(1) 
(the off-diagonal components of the SU(2) gauge field). The disordered phase
played a role similar to the complex scalar field of the GL 
equation. In \cite{dzhsin02a} we simply set this scalar field to a constant 
vacuum expectation value {\it i.e.} we froze the equations of the SU(2)/U(1)
fields making them nondynamical degrees of freedom. In this way we obtained
London's equation. In the present work we want to ``unfreeze'' the
equations connected with the SU(2)/U(1) part, and show that it is possible
to obtain a GL-like equation. This
helps strengthen the dual superconductor picture of the confining 
Yang-Mills vacuum. As a reminder the Lagrangian density of
Ginzburg-Landau is
\begin{equation}
\label{sec1}
{\mathcal L} = - \frac{1}{4}f_{\mu \nu} f^{\mu \nu} + 
   \left( D_\mu \varphi^* \right) \left( D^\mu \varphi \right) - 
   m^2 \left| \varphi \right|^2  - \lambda \left| \varphi \right|^4 
\end{equation}
where $\varphi$ is a complex scalar field, $f_{\mu \nu} =\partial _{\mu}
a_{\nu} -\partial _{\nu} a_{\mu}$, is the field strength tensor of the Abelian
field, $a_{\mu}$, and $D_{\mu} = \partial _{\mu} -i e a_{\mu}$ is the
covariant derivative. 

For the dual superconducting picture it is important that the potential
for the scalar field in \eqref{sec1} be of the spontaneous symmetry breaking
form, and have Nielsen-Olesen flux tube solutions \cite{no}. 
This means that the mass term in \eqref{sec1} should 
be tachyonic ({\it i.e.} $m^2 <0$). In this work the tachyonic mass term 
is generated via the condensation of ghosts as in ref. \cite{dudal} (see
also ref. \cite{lemes}). 

\section{Ordered and disordered phases}

In this section we review the Abelian projection decomposition of 
the  SU(2) gauge field  \cite{kondo} and necessary results from
\cite{dzhsin02a}. The SU(2) gauge fields  $\mathcal{A}_\mu = 
\mathcal{A}^B_\mu T^B$ and $\mathcal{F}^B_{\mu\nu}$ can be 
decomposed as 
\begin{eqnarray}
  \mathcal{A}_\mu & = & \mathcal{A}^B_\mu T^B = a_\mu T^a + A^m_\mu T^m ,
\label{sec2-10a}\\
  a_\mu & \in & U(1) \quad \text{and} \quad A^m_\mu \in SU(2)/U(1)
\label{sec2-10b}\\
    \mathcal{F}^B_{\mu\nu} T^B & = & \mathcal{F}^3_{\mu\nu}T^3 +
  \mathcal{F}^m_{\mu\nu}T^m
\label{sec2-10c}
\end{eqnarray}
where 
\begin{eqnarray}
  \mathcal{F}_{\mu\nu} & = & f_{\mu\nu} + \Phi_{\mu\nu}
  \; \in \; U(1) ,
\label{sec2-20a}\\
  f_{\mu\nu} & = & \partial_\mu a_\nu - \partial_\nu a_\mu 
  \; \in \; U(1) ,
\label{sec2-20b}\\
  \Phi_{\mu\nu} & = & g \epsilon^{3mn} A^m_\mu A^n_\nu \; \in \; U(1) ,
\label{sec2-20c}\\
  \mathcal{F}^m_{\mu\nu} & = & F^m_{\mu\nu} + G^m_{\mu\nu} \; 
  \in SU(2)/U(1) ,
\label{sec2-20d}\\
  F^m_{\mu\nu} & = & \partial_\mu A^m_\nu - \partial_\nu A^m_\mu 
  \; \in \; SU(2)/U(1) ,
\label{sec2-20e}\\
  G^m_{\mu\nu} & = & g \epsilon^{3mn}
  \left(
  A^n_\mu a_\nu - A^n_\nu a_\mu
  \right) \; \in \; SU(2)/U(1)
\end{eqnarray}
where $\epsilon^{ABC}$ are the structural constants of
SU(2), $g$ is the coupling constant, $a=3$ is the index of the Abelian subgroup, and
$m,n = 1,2$ are the indices of the coset. After obtaining the classical
field equations in terms of this decomposition we applied a
quantization technique of Heisenberg \cite{heisenberg} where the classical 
fields were replaced by operators ($a_{\mu} \rightarrow \hat{a}_\mu$
and $A^m_{\mu} \rightarrow \hat{A}^m_\mu$). The 
differential equations for the field operators are
\begin{eqnarray}
  \partial_\nu \left( \hat{f}^{\mu\nu} + \hat{\Phi}^{\mu\nu} \right) & = &
  -g \epsilon^{3mn} \hat A^m_\nu
  \left(
  \hat F^{n\mu\nu} + \hat G^{n\mu\nu}
  \right),
\label{sec2-50a}\\
  D_\nu \left( \hat F^{m\mu\nu} + \hat G^{m\mu\nu} \right) & = &
  -g \epsilon^{3mn}
  \left[
  \hat A^n_\nu \left( \hat f^{\mu\nu} + \hat \Phi^{\mu\nu} \right) -
  \hat a_\nu \left( \hat F^{n\mu\nu} + \hat G^{n\mu\nu} \right)
  \right]
\label{sec2-50b}
\end{eqnarray} 
(We note that in ref. \cite{simonov} similar ideas were 
presented to obtain a set of self coupled equations for the field correlators). 
Equations \eqref{sec2-50a} \eqref{sec2-50b} can be used  to determine 
the physically measurable expectation values for any field operators such as
$\langle Q |\cdots \hat a_\mu(x) \cdots \hat A^m_\nu(y) | Q \rangle$, where
$| Q \rangle$ is some quantum state. As an example, 
if we average equations \eqref{sec2-50a} \eqref{sec2-50b} 
we obtain equations for $\langle \hat A^m_\nu \rangle$ and 
$\langle \hat a_\nu \rangle$. The problem is that the resulting 
differential equations for these expectation values contain additional
terms like $\mathcal{G}^{mn}_{\mu\nu} = \langle A^m_\mu A^n_\nu \rangle$ 
so the system is not closed. Going back to equations \eqref{sec2-50a} \eqref{sec2-50b} 
and differentiating one can obtain an operator equation for the product 
$\hat A^m_\mu \hat A^n_\nu$, which then contains new additional terms.
Continuing in this way one obtains an infinite set of coupled,
differential equations for the various expectation values of field operators
of ever increasing powers. This
process can be paralleled with the standard Feynman diagram procedure
where one has an infinite set of loop diagrams that one must (in
principle) calculate. This infinite system of coupled differential equations
does not have an exact, analytical solution, so one must use some
approximation.
\par
In \cite{dzhsin02a} we used several assumptions that led to a simplification
of this system of differential equations. The two main assumptions were 
\begin{enumerate}
  \item
  After quantization the components $\hat A^m_\mu (x)$
  become stochastic. In mathematical terms this assumption means
\begin{equation}
  \left\langle A^m_\mu (x) \right\rangle = 0
  \qquad \text{and} \qquad
  \left\langle
  A^m_\mu (x) A^n_\nu (x) \right\rangle \neq 0 
  \label{sec2-60}
\end{equation}
Later we will give a specific form for the nonzero term. 
   \item
  The gauge potentials $a_\mu$ and $A^m_\mu$ are not correlated, and
  $a_{\mu}$ behaves in a classical manner.
  Mathematically this means that
\begin{equation}
  \left\langle f(a_\mu) g(A^m_\nu) \right\rangle =
  \left\langle f(a_\mu) \right\rangle
  \left\langle g(A^m_\mu) \right\rangle =
  f(a_\mu)  \left\langle g(A^m_\mu) \right\rangle 
\label{sec2-80}
\end{equation}
  where $f,g$ are any functions of $a _{\mu}$ and $A^m _{\mu}$ respectively.
  The classical behavior of $a _{\mu}$  results in $\left\langle f(a_\mu) \right\rangle
   \rightarrow  f(a_\mu)$. 
\end{enumerate}
One way in which the present work will deviate from the procedure of ref. \cite{dzhsin02a} is
that rather than working in terms of the equations of motion of the SU(2) gauge theory, we
will focus on the Lagrange density.

\section{Ginzburg - Landau Lagrangian}

We now want to show that an effective, complex, Higgs-like, scalar 
field can be obtained from the SU(2)/U(1) coset part of the SU(2) 
gauge theory. The self interaction of this effective scalar field is 
a consequence of nonlinear terms in the original Yang-Mills Lagrangian. 
The mass term for the scalar field is generated via the 
condensation of ghosts fields as discussed in the following section.

Making a connection between scalar and gauge fields is not a new idea. In
ref. \cite{corr} it was shown that by setting a non-Abelian gauge field to 
some combination of a scalar field and its
derivatives it was possible to obtain massless
$\lambda \varphi ^4$ theory. One could
obtain a massive $\lambda \varphi ^4$ theory by starting with
Yang-Mills theory with a mass term inserted by hand \cite{actor}. 
The final scalar field Lagrangian that we will arrive at is also
a massless $\lambda \varphi ^4$ theory with the addition of a 
coupling to a U(1) gauge field. This is (except for the U(1) gauge
field coupling) similar to the result of \cite{corr}. In the present
paper we exchange the two gauge field of the SU(2)/U(1) coset 
with a complex scalar field, whereas in Refs. \cite{corr} \cite{actor}
one gauge field is exchanged for a real scalar field. We begin by taking 
the expectation of the SU(2) Lagrangian (the gauge fixing and Faddeev-Popov 
terms are dealt with in the next section)
\begin{equation}
  -\frac{1}{4}\left\langle Q \left| 
  \cal F^{\cal A}_{\mu \nu} \cal F^{\cal A \mu \nu}  
  \right | Q \right\rangle = 
  -\frac{1}{4} \left\langle 
  \cal F^{\cal A}_{\mu \nu} \cal F^{\cal A \mu \nu} 
  \right\rangle = 
  -\frac{1}{4} \left\langle 
  {\cal F}^3_{\mu \nu} {\cal F}^3_{\mu \nu} + 
  {\cal F}^m_{\mu \nu} {\cal F}^m_{\mu \nu} 
  \right\rangle 
\label{sec3-10}
\end{equation}
This contains many terms which we will now consider in order. 
First we consider the term 
\begin{equation}
  \left\langle 
  {\cal F}^3_{\mu \nu} {\cal F}^3_{\cal A \mu \nu} 
  \right\rangle = 
  \left\langle
    \left(
    f_{\mu \nu} + \Phi_{\mu \nu}
    \right)
    \left(
    f^{\mu \nu} + \Phi^{\mu \nu}
    \right)
  \right\rangle
\label{sec3-20}
\end{equation}
where the field $f_{\mu \nu}$ is in the ordered phase and $\Phi_{\mu \nu}$ 
is in the disordered phase. From the second assumption of the previous
section  $a_\mu$ and 
$f_{\mu \nu} = \partial_\mu a_\nu - \partial_\nu a_\mu$, behave
as classical fields so that
\begin{eqnarray}
  \left\langle f_{\mu \nu} \Phi^{\mu \nu} \right\rangle & = & 
  f_{\mu \nu} \left\langle \Phi^{\mu \nu} \right\rangle 
\label{sec3-30a}\\
  \left\langle f_{\mu \nu} f^{\mu \nu} \right\rangle & = & 
  f_{\mu \nu}  f^{\mu \nu}   
\label{sec3-30b}\\
  \left\langle \Phi_{\mu \nu} \right\rangle & = & 
  g \epsilon^{3mn} \left\langle A^m_\mu A^n_\nu \right\rangle  
\label{sec3-30c}  
\end{eqnarray} 
For the expectation of  $A^m_\mu(y) A^n_\nu(x) $ we take the form
\begin{equation}
  \left\langle A^m_\mu(y) A^n_\nu(x) \right\rangle = 
  - \delta^{mn} \eta_{\mu \nu} {\cal G}(y,x)
\label{sec3-40}
\end{equation}
${\cal G}(y,x)$ is an arbitrary function. This is a general 
form consistent with the color and Lorentz indices of the left 
hand side. One might think to add a term with an index structure like
 $\eta_{\mu \nu} \epsilon^{3 mn}$. However, this is antisymmetric under exchange
of $A^m_\mu(y) A^n_\nu(x)$ ({\it i.e.} exchanging both Lorentz and group indices)
which is not consistent with the bosonic statistics of the gauge fields. 
The quantity in \eqref{sec3-40} is a mass dimension 2 condensate. The role of
such gauge non-invariant quantites in the Yang-Mills vacuum has been discussed by
several authors \cite{kondo1} \cite{gubarev}. 

From \eqref{sec3-40} we find 
\begin{equation}
  \left\langle \Phi_{\mu \nu} \right\rangle = 0
\label{sec3-50}
\end{equation}
\begin{equation}
  \left\langle
  \Phi_{\mu\nu} \Phi^{\mu\nu}
  \right\rangle =
  2 \left\langle
  A^1_\mu A^2_\nu A^{1\mu} A^{2\nu}
  \right\rangle -
  2 \left\langle
  A^2_\mu A^1_\nu A^{1\mu} A^{2\nu}
  \right\rangle.
\label{sec3-60}
\end{equation}
We now assume that 
\begin{equation}
  \left\langle
  A^m_\mu(x_1) A^n_\nu(x_2) A^{p\mu}(x_3) A^{q\nu}(x_4) 
  \right\rangle \approx
  \left\langle
  A^m_\mu(x_1) A^{p \mu}(x_3)
  \right\rangle
  \left\langle
  A^n_\nu(x_2) A^{q \nu}(x_4)
  \right\rangle
\label{sec3-70}
\end{equation}
which gives
\begin{equation}
  \left\langle
  \Phi_{\mu\nu}(x) \Phi^{\mu\nu}(x) 
  \right\rangle \approx
  32 g^2 {\cal G}^2 (x,x)
\label{sec3-80}
\end{equation} 
The next term is
\begin{equation}
\begin{split}
  \left\langle
  \mathcal {F}^m_{\mu \nu} (y)
  \mathcal {F}^{m \mu \nu} (x)
  \right\rangle \Bigr |_{y=x} =
  \left(
    \left[
    \partial_{\mu y}A^m_\nu (y) -  g a_\mu (y)
      \epsilon^{3mn} A^n_\nu (y)
    \right] -
    \left[
    \partial_{\nu y}A^m_\mu (y) - g a_\nu (y)
      \epsilon^{3mn} A^n_\mu (y)
    \right]
  \right)
\\
  \left(
    \left[
    \partial^\mu_x A^{m \nu} (x) - g a^\mu (x)      
      \epsilon^{3mp} A^{p \nu} (x)
    \right] -
    \left[
    \partial^\nu_x A^{m \mu} (x) - g a^\nu (x)
      \epsilon^{3mp} A^{p \mu} (x)
    \right]
  \right) \Bigr |_{y=x} .
\end{split}
\label{sec3-90}
\end{equation}
We begin by expanding the first term of $\mathcal {F}^m_{\mu \nu} (y)$
in square brackets against the first term of $\mathcal {F}^{m \mu \nu} (x)$.
This has four terms the first of which is
\begin{equation}
  \left\langle
  \partial_{\mu y}A^m_\nu (y)
  \partial^\mu_xA^{m \nu} (x)
  \right\rangle \Bigr |_{y=x} =
  \partial_{\mu y} \partial^\mu_x
  \left\langle
  A^m_\nu (y) A^{m \nu} (x)
  \right\rangle \Bigr |_{y=x} =
  - \partial_{\mu y} \partial^\mu_x \delta^{mm} \eta^\nu_\nu
  \mathcal G (y,x) \Bigr |_{y=x} =
  - 8 \partial_{\mu y} \partial^\mu_x \mathcal G (y,x) \Bigr |_{y=x} 
\label{sec3-100}
\end{equation}
The next two terms are
\begin{eqnarray}
 g^2  \epsilon^{3mn} \epsilon^{3mp} a_\mu (x) a^\mu (x)
  \left\langle
  A^n_\nu (x) A^{p \nu} (x)
  \right\rangle & = & 
  -8 g^2 a_\mu a^\mu (x) \mathcal G (x,x)
\label{sec3-110}\\
  \left\langle
    \left[
    \partial_{\mu y} A^m_\nu (y) 
    \right]
    \left[
    g \epsilon^{3mn} a^\mu (x)
    A^{n \nu}(x)
    \right]
  \right\rangle \Bigr |_{y=x} & = & 
  g \epsilon^{3mn} a^\mu (x)
  \left\langle
    \left[
    \partial_{\mu y} A^m_\nu (y)
    \right]
  A^{n \nu}(x)
  \right\rangle \Bigr |_{y=x} 
\label{sec3-120}
\end{eqnarray}
We want to consider this last term more closely. From 
eqs. \eqref{sec3-40}, \eqref{sec3-100} and \eqref{sec3-110} 
it can be seen that there are quantum 
correlations only between fields with the same color ($\delta^{mn}$) and 
coordinate ($\eta_{\mu \nu}$) indices. In the absence of some principle
which forbids it, one would in general expect that there should physically be some
interaction between gauge fields of different colors. In eq. \eqref{sec3-40} we
excluded terms proportional to $\eta _{\mu \nu} \epsilon^{3mn}$ because of
the Bose symmetry of the gauge fields. In the present expression the
gauge fields do not appear symmetrically (one is acted on by a derivative operator)
so such a term can be included. Thus we take the expectation of \eqref{sec3-120}
to have the general form
\begin{equation}
  \left\langle
    \left[
    \partial_{\mu y} A^m_\alpha (y) 
    \right]
  A^n_\beta
  \right\rangle = 
  - \delta^{mn} \eta_{\alpha \beta} \partial_{\mu y} 
  \mathcal G (y,x) - i \epsilon^{3mn} \eta_{\alpha \beta}
  \partial_{\mu y} \mathcal P (y,x)  .
\label{sec3-130}
\end{equation}
where $\mathcal P (y,x)$ is some general function. This new term will
mix gauge bosons of different colors. Using this form eq. \eqref{sec3-120}
becomes
\begin{equation}
  \left\langle 
    \left[
    \partial_{\mu y} A^m_\nu (y) 
    \right]
    \left[
    g \epsilon^{3mn} a^\mu (x)
    A^{n \nu}(x)
    \right]
  \right\rangle \Bigr |_{y=x}  = 
  - 8 i g a^\mu(x) \partial_{\mu y} \mathcal P (y,x) \Bigr |_{y=x} .
\label{sec3-150}
\end{equation}
Making the same assumption for the other cross term 
\begin{equation}
  \left\langle
  A^n_\alpha (y) 
    \left[
    \partial_{\mu x} A^m_\beta (x)
    \right]
  \right\rangle = 
  -\delta^{nm} \eta_{\alpha \beta} \partial_{\mu x} \mathcal G (y,x) - 
  i \epsilon^{3nm} \eta_{\alpha \beta} \partial_{\mu x} \mathcal P^* (y,x) .
\label{sec3-160}
\end{equation}
yields
\begin{equation}
  \left\langle
  A^n_\nu (y) 
    \left[
    g \epsilon^{3mn} a_\mu (x) \partial^\mu_x A^{m\nu} (x)
    \right]   
  \right\rangle = 
  8 i g a^\mu (x) \partial_{\mu x} \mathcal P^* (y,x) \Bigr |_{y=x} 
\label{sec3-170}
\end{equation}
Next we expand the first term of $\mathcal {F}^m_{\mu \nu} (y)$
in square brackets against the second term of $\mathcal {F}^{m \mu \nu} (x)$.
This also has four terms which are
\begin{eqnarray}
  \left\langle
    \left[
    \partial_{\mu y} A^m_\nu (y) 
    \right]
    \left[
    \partial^\nu_x A^{m\mu} (x) 
    \right]
  \right\rangle \Bigr |_{y=x} & = & 
  -2 \partial_{\mu y} \partial^\mu_x \mathcal G (y,x) \Bigr |_{y=x} , 
\label{sec3-180}\\
  \left\langle
    \left[
    \partial_{\mu y} A^m_\nu (y) 
    \right]
    \left[
    g a^\nu (x) \epsilon^{3mp} A^{p\mu} (x) 
    \right]
  \right\rangle \Bigr |_{y=x} & = & 
  -2 i g a^\nu (x) \partial_{\nu y}   \mathcal P (y,x) \Bigr |_{y=x} , 
\label{sec3-180a}\\
  \left\langle
    \left[
    g a_\mu (y) \epsilon^{3mn} A^n_\nu (y) 
    \right] 
    \left[
    \partial^\nu_x A^{m\mu} (x) 
    \right]
  \right\rangle & = & 
  2 i g a^\mu (x) \partial_{\mu x} \mathcal P^* (y,x) \Bigr |_{y=x} , 
\label{sec3-180b}\\
  \left\langle
    \left[
    g a^\mu (y) \epsilon^{3mn} A^n_\nu (y) 
    \right] 
    \left[
    g a^\nu (x) \epsilon^{3mp} A^{p\mu} (x) 
    \right] 
  \right\rangle & = & 
  - 2 g^2 a_\mu (x) a^\mu (x) \mathcal G (x,x) . 
\label{sec3-180c}
\end{eqnarray}  
Finally we need to expand the second term of  $\mathcal {F}^m_{\mu \nu} (y)$
against the second term of $\mathcal {F}^{m \mu \nu} (x)$, and
also the second term of $\mathcal {F}^m_{\mu \nu} (y)$ against the first term 
of $\mathcal {F}^{m \mu \nu} (x)$. These yield, respectively, the same results
as eqs. \eqref{sec3-100}, \eqref{sec3-110},
\eqref{sec3-150}, and \eqref{sec3-170}, and 
eqs. \eqref{sec3-180}-\eqref{sec3-180c} . Collecting all these terms together
gives
\begin{equation}
\label{sec3-185}
 \left\langle
  \mathcal F^m_{\mu \nu} \mathcal F^{m \mu \nu}
  \right\rangle = 
-20 \left[ \partial_{\mu y} \partial^\mu_x \mathcal G (y,x)
             + g^2 a_\mu (x) a^\mu (x) \mathcal G (x,x)
 -  i g a^\mu (x) \partial_{\mu x} \mathcal P^* (y,x) 
+ i g a^\mu (x) \partial_{\mu x} \mathcal P (y,x) \right] \Bigr |_{y=x} 
\end{equation}
We now make the
approximation that both functions, $\mathcal G (y,x)$ and $\mathcal P (y,x)$, 
can be rewritten in terms of a single complex scalar function as
\begin{equation}
  \mathcal G (y,x) = \mathcal P(y,x) = 
  \frac{1}{5} \varphi^*(y) \varphi(x) 
\label{sec3-190}
\end{equation}
Setting both $\mathcal G (y,x) $ and  $\mathcal P(y,x)$ equal
to the same product of a complex scalar field  is called the 
\textit{one function approximation or ansatz}. The factor of $1/5$
is to ensure that the kinetic term of this scalar field will have a
factor of $1$ in front of it. Taking eqs.
\eqref{sec3-40}, \eqref{sec3-130}, \eqref{sec3-190} together, we
want to note the similarity between this ansatz for the gauge fields, and the
ansatz used in ref. \cite{corr}. Both have similar forms for the indexed
terms, and both involve some form of scalar field and its derivatives.  
This complex function, $\varphi$,  will act as the effective scalar field.
In ref. \cite{dzhsin02a} we set $\varphi^* \varphi = const.$, making the 
degrees of freedom connected with the SU(2)/U(1) coset space nondynamical.
(Note that in ref. \cite{dzhsin02a} we used a single scalar field. The scalar
field of ref. \cite{dzhsin02a} is proportional to $\varphi^*(x) \varphi(x) $
in the present paper). By letting $\varphi$ be a function of the coordinates
we make these dynamical degrees of freedom.
Using \eqref{sec3-190} in \eqref{sec3-185} we find
\begin{equation}
  \left\langle
  \mathcal F^m_{\mu \nu} \mathcal F^{m \mu \nu}
  \right\rangle = 
  -4\left[
  \left( \partial^\mu \varphi^* \right)
  \left( \partial_\mu \varphi \right) + g^2 a^2_\mu \varphi^* \varphi - 
  i g a^\mu \left( \partial_\mu \varphi^* \right) \varphi + 
  i g a^\mu \varphi^* \left( \partial^\mu \varphi \right)
  \right] = 
  -4\left| \partial_\mu \varphi - i ga_\mu \varphi \right|^2 .
\label{sec3-200}
\end{equation}
Thus the total Lagrangian density is 
\begin{equation}
  \left \langle {\mathcal L} \right\rangle = 
  - \frac{1}{4} \left \langle 
  \mathcal {F^A}_{\mu \nu} \mathcal F^{\mathcal A \mu \nu} 
   \right\rangle = - \frac{1}{4}f_{\mu \nu} f^{\mu \nu} + 
    \left( D_\mu \varphi^* \right) \left( D^\mu \varphi \right) - 
   \frac{8 g^2}{25} \left| \varphi \right|^4 
\label{sec3-210}
\end{equation}
where $D_\mu = \partial_\mu - i g a_\mu$. This is the GL
Lagrangian,  with a {\it massless}, effective scalar field. This
scalar field is connected with the off diagonal gauge fields by 
\eqref{sec3-40} \eqref{sec3-190}. This lack of a mass
term is a shortcoming for the effective Lagrangian  of  \eqref{sec3-210}.
Without it there is no spontaneous symmetry breaking and
no Nielsen-Olesen flux tube solutions, both which are critical to
make the connection to the dual superconducting picture of the QCD vacuum.
In the next section we show how a condensation of ghosts fields can
lead to a mass term for the effective scalar field $\varphi$.
This mass term is of the correct form ({\it i.e.} tachyonic) to give rise to
spontaneous symmetry breaking and Nielsen-Olesen flux tube solutions. 

\section{Tachyonic mass term via ghost condensation}

In ref. \cite{dudal} (see also \cite{lemes}) it was shown that a tachyonic mass term is 
generated for the off-diagonal gauge fields of a pure SU(2) Yang-Mills via a condensation of
ghost and anti-ghost fields. This result was taken to be somewhat of a problem
since this would apparently give the off diagonal gluons a tachyonic mass.
In the present work having a tachyonic mass for our effective scalar field
(which is related to the off diagonal gauge fields via \eqref{sec3-40} 
\eqref{sec3-190}) is a desired result.
We sketch the parts of ref. \cite{dudal} relevant to the present
work. To take care of gauge fixing and integration over gauge equivalent 
field configurations one must add a gauge fixing (${\mathcal L} _{GF}$) and  
Faddeev-Popov part (${\mathcal L} _{FP}$) to the Yang-Mills Lagrangian in
\eqref{sec3-210}
\begin{equation}
\label{sec4-10}
{\mathcal L}_{GF} + {\mathcal L}_{FP} = i \delta _B {\overline{ \delta}}_B \left(
\frac{1}{2} A_{\mu} ^m A^{\mu m} - \frac{\alpha}{2} i C^m {\overline{ C}}^m \right)
\end{equation}
where $\alpha$ is a gauge parameter, $C^m$ and ${\overline{ C}}^m$ are the off diagonal ghost and
anti-ghost field, and $\delta _B$ and ${\overline {\delta}}_B$ are the BRST and anti-BRST
transformations. These gauge fixing and Faddeev-Popov Lagrangians can be
transformed into
\begin{equation}
\label{sec4-20}
{\mathcal L}_{GF} + {\mathcal L}_{FP} = -\frac{1}{2 \alpha} (D_{\mu} ^{mn} A^{\mu n} )^2
+ i {\overline{ C}}^m D^{mp} _{\mu} D^{\mu pn} C^n - ig^2 \epsilon ^{mq} \epsilon ^{pn} {\overline{ C}}^m C^n A^{\mu p}
A_{\mu} ^q + \frac{\alpha}{4} g^2 \epsilon ^{mn} \epsilon ^{pq} {\overline{ C}}^m {\overline{ C}}^n C^p C^q
\end{equation}
where $\epsilon ^{mn}$ is the antisymmetric symbol for the off-diagonal indices ($\epsilon ^{12} =
-\epsilon ^{21} =1$ and $\epsilon ^{11}=\epsilon ^{22}=0$), and $D_{\mu} ^{mn} =
\partial _{\mu} \delta ^{mn} - g \epsilon ^{mn} a_{\mu}$.

Next, the last term in \eqref{sec4-20} is a four ghost interaction term which can be replaced by
\cite{dudal} \cite{lemes}
\begin{equation}
\label{sec4-30}
\frac{\alpha}{4} g^2 \epsilon ^{mn} \epsilon ^{pq} {\overline{ C}}^m {\overline{ C}}^n C^p C^q
\rightarrow -\frac{1}{2 \alpha g^2} \psi ^2 - i \psi \epsilon ^{mn} {\overline{ C}}^m C^n
\end{equation}
with $\psi$ being an auxiliary field. Extracting the ghost kinetic energy term from the
second term in \eqref{sec4-20} and adding it to \eqref{sec4-30} one arrives at a
subset of the Lagrangian involving only the ghost and auxiliary field
\begin{equation}
\label{sec4-40}
{\mathcal L}_{ghost} = i {\overline{ C}}^m \partial _{\mu} \partial ^{\mu} C^m
-\frac{1}{2 \alpha g^2} \psi ^2 - i \psi \epsilon ^{mn} {\overline{ C}}^m C^n
\end{equation}
The Coleman-Weinberg mechnaism \cite{coleman} can be applied to 
${\mathcal L}_{ghost}$ leading to a spontaneous symmetry breaking
potential for $\psi$ of the form
\begin{equation}
\label{sec4-50}
V_{eff} (\psi ) = \frac{\psi ^2}{2 \alpha {\overline g}^2} + \frac{\psi ^2}{32 \pi ^2}
\left( \ln \frac{\psi ^2}{{\overline \mu} ^4} -3 \right)
\end{equation}
This potential has a non-zero minimum at $\psi = \pm v = \pm {\overline \mu}^2 \exp \left(1 - 
\frac{8 \pi}{\alpha {\overline g}^2 {\overline \mu}^2} \right)$ (the bars over the quantities
indicate that the $\overline{MS}$ dimensional regularization scheme is being employed).
Investigating the ghost propagator in this non-zero vacuum it is found \cite{dudal} 
that a ghost condensation occurs
\begin{equation}
\label{sec4-60}
\langle i {\overline{ C}} ^m C^m \rangle = -\frac{v}{16 \pi} <0
\end{equation}
since $v>0$. The third term in the Lagrangian in \eqref{sec4-20} now becomes
\begin{equation}
\label{sec4-70}
ig^2 \epsilon ^{mq} \epsilon^{pn} {\overline{ C}}^m C^n A^{\mu p}A_{\mu} ^q
\rightarrow
\frac{1}{2} g^2 \langle i {\overline{ C}} ^m C^m \rangle \langle A_{\mu} ^n A^{\mu n} \rangle
=
\frac{1}{2} g^2 \left( - \frac{v}{16 \pi} \right) \left( -8 {\cal G} \right)
=
\frac{v g^2}{20 \pi} \varphi ^* \varphi
\end{equation}
where \eqref{sec3-40}  and \eqref{sec3-190} have been used.
Putting this term together with the quartic term from the effective GL Lagrangian in
\eqref{sec3-210} we find that $\varphi$ has developed an effective potential
of the form
\begin{equation}
\label{sec4-80}
V_{\varphi} = -\frac{v g^2}{20 \pi} | \varphi |^2 + \frac{8 g^2}{25} |\varphi |^4
\end{equation}
A tachyonic mass term ($m^2 = -(v g^2)/(20 \pi) <0$)  has been generated 
for the effective scalar field, $\varphi$, with the consequence that spontaneous
symmetry breaking occurs. We have arrived at an effective
{\it massive} GL-Lagrangian with the tachyonic mass term for the $\varphi$
field  being generated by ghost condensation. This effective GL-Lagrangian will 
also have Nielsen-Olesen flux tube solutions.

With this tachyonic mass term for $\varphi$ spontaneous symmetry
breaking occurs and the $U(1)$ field, $a^{\mu}$, will develop a mass of $\sqrt{ (5g^2 v)/(32 \pi )}$.
It might be thought that the ghost condensate would also contribute some to the mass
for $a^{\mu}$, since the second term in the gauge fixing and Faddeev-Popov Lagrangian
of \eqref{sec4-20} contains a term proportional to $i {\overline{ C}}^a C^a a_{\mu} a^{\mu}$.
However, in \cite{dudal} it was argued that this term is canceled by 1-loop contributions
coming from other terms in \eqref{sec3-210} which are proportional to $\epsilon ^{mn}
\left( \partial _{\mu} {\overline{ C}} ^m \right) C^n a^{\mu}$ and $\epsilon ^{mn}
{\overline{ C}} ^m \left( \partial _{\mu} C^n \right) C^n a^{\mu}$. Thus the mass of $a^{\mu}$
is generated purely from the spontaneous symmetry breaking of the effective GL
Lagrangian \eqref{sec3-210}. 

\section{Conclusions}

In this paper we have combined several ideas (Abelian projection, quantization 
methods originally proposed by Heisenberg, and some assumptions about the
forms of various expectation values of the gauge fields) to show that
one can construct an effective scalar field within a pure SU(2) gauge field 
theory. The system of effective scalar field plus the remaining Abelian 
field is essentially scalar electrodynamics or the relativisitic version
of the GL Lagrangian. There is also a
connection between the present work and Cho's \cite{cho} ``magnetic
symmetry'' study of the dual Meissner effect within Yang-Mills theory.
In Refs. \cite{cho} a Lagrangian similar to our eq. \eqref{sec3-210}
is obtained, with the complex scalar field being associated with a monopole
coupled to a U(1) dual magnetic gauge boson. This may offer one possible
physical interpretation of our scalar field of eq. \eqref{sec3-190}:
it may represent some effective monopole-like field which results from
the SU(2)/U(1) coset fields. This is in accord with lattice QCD simulations
which indicate that monopole condensation plays a role in
color confinement \cite{kron}.  

The effective mass term for the Lagrangian in \eqref{sec3-190} comes from
a condensation of ghost fields. This condensation is of the correct character
({\it i.e.} tachyonic) to give spontaneous
symmetry breaking and the existence of Nielsen-Olesen flux tube solutions 
\cite{no}. Both of these are thought to be important features in 
explaining confinement via the dual superconducting model of the QCD vacuum.
The apparently problematic result ({\it i.e.} the tachyonic ghost condensate and
resulting tachyonic nature for the off diagonal gauge field masses)
of refs. \cite{dudal} \cite{lemes} is actually a desired result in the
present work.In our model a tachyonic mass
term is neccessary for the effective field, $\varphi$ in order to have
a GL-like Lagrangian that exhibits both spontaneous symmetry breaking and
Nielsen-Olesen flux tube solutions. Note that 
$\varphi$ is related to the condensate of the off-diagonal gauge fields
via \eqref{sec3-40} and \eqref{sec3-190}. 

The essential physics here is that one has disordered fields (the gauge fields
of the SU(2)/U(1) coset space or equivalently the effective, complex, scalar field)
which ``pushes out'' ({\it i.e.} exhibits the Meissner effect)
the ordered field (the Abelian, U(1) field) except in the interior of the flux tubes.
This is a continuation of ref. \cite{dzhsin02a} which supports the dual superconducting
picture of the Yang-Mills vacuum for a {\it pure} gauge field. The scalar field
comes from some subset of the gauge fields rather than being put in by
hand. An interesting and open question is
if the procedure in this paper can be applied to the SU(3) gauge theory of the
strong interaction. In ref. \cite{par} the SU(3) Lagrangian {\it with quarks} was
studied, and using a procedure similar to the one in the  present paper, it was found that
the dual Meissner effect did occur.
\par 
The main assumptions used to arrive at the effective GL-like
Lagrangian from a pure non-Abelian gauge theory are enumerated as follows
\begin{itemize}
    \item 
    SU(2) quantized gauge fields can be separated on two components : 
    ordered (Abelian) and disordered (coset) phases; 
    \item 
    ordered and disordered fields are correlated according to eq. \eqref{sec2-80}; 
    \item 
    Green's functions for the disordered phase can be expressed via 
    one function approximation as in eq. \eqref{sec3-190}; 
      \end{itemize}    
In concluding we would like to make the following remark: In this paper 
we have shown that in non-linear quantum systems one can have  
well-ordered objects (Nielsen - Olesen flux tube in our case). This can be 
compared with classical self-organizing systems like Benard cells \cite{ben}, or
Belousov-Zhabotinsky \cite{jab} reactions (and references therein). 
The most important difference is that in 
the classical case the self-organization arises from some external flux 
of energy, but in our quantum case the self-organization arises from inside the 
non-linear, quantum system without the need for any external sources.

\section{Acknowledgment}

VD is grateful Prof. H. Kleinert for invitation for the research 
and discussion, Prof. V. Gurovich for the discussion, 
DAAD for the financial support and Alexander von Humboldt Foundation 
for the support of this work.

\end{document}